# A Review of Localization and Tracking Algorithms in Wireless Sensor Networks

Sudhir Kumar
Department of Electronics and Communication Engineering
Visvesvaraya National Institute of Technology, Nagpur, India
sudhirk.iitk@gmail.com

Rajesh M. Hegde
Department of Electrical Engineering
Indian Institute of Technology, Kanpur, India
rhegde@iitk.ac.in

*Abstract*—In this paper, a comprehensive survey of the pioneer as well as the state of-the-art localization and tracking methods in the wireless sensor networks is presented. Localization is mostly applicable for the static sensor nodes, whereas, tracking for the mobile sensor nodes. The localization algorithms are broadly classified as range-based and range-free methods. The estimated range (distance) between an anchor and an unknown node is highly erroneous in an indoor scenario. This limitation can be handled up to a large extent by employing a large number of existing access points (APs) in the range free localization method. Recent works emphasize on the use multi-sensor data like magnetic, inertial, compass, gyroscope, ultrasound, infrared, visual and/or odometer to improve the localization accuracy further. Additionally, tracking method does the future prediction of location based on the past location history. A smooth trajectory is noted even if some of the received measurements are erroneous. Real experimental set-ups such as National Instruments (NI) wireless sensor nodes, Crossbow motes and hand-held devices for carrying out the localization and tracking are also highlighted herein.

## I. Introduction

Localization and tracking of wireless sensor nodes is a challenging issue in a large mobile sensor network. This assumes importance in several applications including location-based services, network optimization, and environment characterization [1, 124]. Localization methods are used to compute the position of the node in a 2D/3D space. The primary applications of the localization methods include sensing, monitoring, surveillance and tracking. For example, military surveillance, intruder detection, environment monitoring (e.g. temperature, humidity and soil health), agriculture monitoring, health monitoring (e.g. smart home), habitat monitoring, pollution control and space application on planet and Moon. It can be utilized for terrestrial, underwater or underground environments. The various research area in wireless sensor networks are connectivity and coverage, localization and tracking, data aggregation and latency minimization, privacy and security, time synchronization, MAC protocol and sleep scheduling, cross layer optimization, routing and topology control, and congestion control. The location of the sensor nodes are used for solving higher layer protocol problems like routing, and data aggregation. Hence, it is of sufficient interest to survey the localization and tracking methods in wireless sensor networks, in this work. In general, these methods utilize the knowledge of the locations of a small subset of nodes which are called anchors [2]. The node that is aware of its location is known as anchor, beacon, source, or landmark in literature. Localization is a one-time process for static nodes, whereas tracking is the continuous localization of the mobile node over time. Tracking also estimates velocity at each instant. The past location information is also incorporated into the tracking method for improved accuracy and for obtaining smoothed trajectories. A broad classification of localization methods is discussed in the ensuing section.

## II. Broad Classification of Localization Methods

Localization methods can be broadly classified as range-free and range based methods in literature [124]. The class tree is illustrated in Figure 1. Range based localization methods utilize the distance between the anchor and unknown node. On the other hand, range-free methods utilize only connectivity information. Range-based methods are fine-grained localization methods, whereas, range-free methods are coarse-grained. A particular method can be chosen based on whether course grained resolution (± 5 m) or fine grained resolution (± 10 cm) is required.

### A. Range-based Localization Methods

Range based methods [128, 130, 131, 132] can be classified into four major classes Received Signal Strength (RSS), Angle-of-Arrival (AOA), Time-of-Arrival (TOA), and Time-Difference-of-Arrival (TDOA). The range based methods like RSS [3, 4, 5, 6, 7, 8], AOA [9, 10, 11, 12, 13], TOA [14, 15, 16, 17, 18], TDOA [19, 15, 20, 21, 22] are extensively discussed in literature. Sensor node localization using RSS provides a cost-effective solution. RSS-based positioning system using visible light communication and accelerometer is proposed in [23] using LOS model. Localization method using TOA and TDOA requires highly precise synchronization [15]. DOA-based methods require a special hardware to estimate the angle at which the signal arrives at the antenna array [9]. The method proposed in [17, 18] describes the node localization scheme based on time of arrival (TOA) of signal.

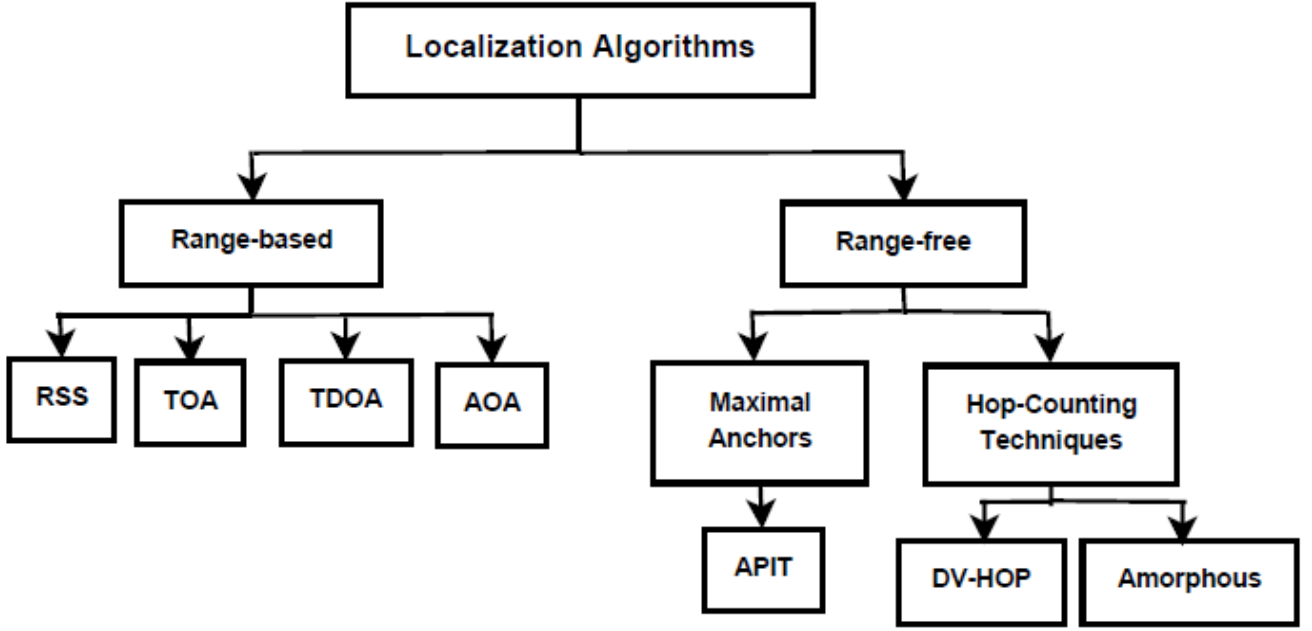

Figure 1: Classification tree for range-based and range-free methods for localization

Time of arrival (TOA) method performs localization using the information about time of arrival of signal from different beacons. But TOA requires a reference time stamp to synchronize the local clock of the node. To overcome this issue of reference offset, a time difference of arrival (TDOA) method has been proposed in [21, 22]. TDOA is based on the principle of time difference of arrival of the beacon signals at a pair of nodes. In this method, one of the beacons is generally taken as a reference sensor. The TDOA method has high localization accuracy in low noise environments. However, this method is highly susceptible to errors from non-line of sight (NLOS) measurements. In [17, 18], TOA-based localization under NLOS scenario is discussed. Additionally, these methods do not assume the statistics of NLOS distribution.

*a) Time-of-Arrival (TOA)- based Localization*

The TOA-based localization method [24] is formulated as

$$g_i(\mathbf{X}) = v(t_i - \tau) - \sqrt{(x_i - x)^2 + (y_i - y)^2}, \quad i \in \mathcal{N} \quad (1)$$

Where, X = x y τ T is the unknown parameter vector. v and N denote the speed of signal and the set of unknown nodes. The Least-Square estimate of location is expressed as

$$G(\mathbf{X}) = \sum_{i \in \mathcal{N}} k_i g_i^2(\mathbf{X}) \quad (2)$$

where $k$ is the confidence parameter between 0 and 1. The location of unknown sensor node can be estimated as

$$\tilde{\mathbf{X}} = \underset{\mathbf{X}}{\mathrm{argmin}} \; G(\mathbf{X}) \quad (3)$$

TOA-based method require synchronization between the node and the anchor and it utilizes distance information between node and anchor

*b) Time-Difference-of-Arrival (TDOA)- based Localization*

TDOA-based localization method [25] can be formulated as follows: The received signals at two different nodes are given by

$$\begin{aligned} y_1(t) &= \alpha_1(t)x(t - \tau_1) + w_1(t) \\ y_2(t) &= \alpha_2(t)x(t - \tau_2) + w_2(t) \end{aligned} \quad (4)$$

where $\alpha_i$ represents the attenuation factor for the anchor and $i^{th}$ node, and $w_i$ is zero-mean Gaussian noise. $\tau_i$ denotes the delay associated with the $i^{th}$ node. Estimated cross-correlation is expressed as

$$\tilde{R}_{y_1 y_2}(\tau) = \frac{1}{T} \int_0^T y_1(t) y_2(t - \tau) \, dt \quad (5)$$

where, $T$ is the observation period. The TDOA can be estimated by maximizing the correlation function as

$$\mathrm{TDOA} = \underset{\tau}{\mathrm{argmax}} \; \tilde{R}_{y_1 y_2}(\tau) \quad (6)$$

Difference between TOAs of pair of nodes that is TDOA is utilized herein for obtaining the location of unknown sensor node. It requires highly precise synchronization between nodes, however, it does not require synchronization between node and source. It may be noted that TDOA is an efficient estimator compared to the TOA-based method.

*c) Angle-of-Arrival (AOA)- based Localization*

The angle and location relationship between the unknown sensor and anchor can be expressed as [10]

$$(x_i - x) = (y_i - y) \tan \psi, \quad i \in \mathcal{N} \quad (7)$$

where, $X = [x\ y]^T$ is an unknown node, and $X_i = [x_i\ y_i]^T$, $i^{th}$ is the anchor. The location of an unknown sensor node using Least-Square approach can be estimated as

$$\tilde{X} = (A^T A)^{-1} A^T B \quad (8)$$

where,

$$A = \begin{bmatrix} 1 & -\tan \psi_1 \\ 1 & -\tan \psi_2 \\ \vdots & \vdots \\ 1 & -\tan \psi_N \end{bmatrix} \quad B = \begin{bmatrix} x_1 - y_1 \tan \alpha_1 \\ x_2 - y_2 \tan \alpha_2 \\ \vdots \\ x_N - y_N \tan \alpha_N \end{bmatrix}$$

It may be noted that AOA-based method uses angle information and it is having high cost because of use of antenna array.

*d) Received Signal Strength (RSS) based Localization*

The RSS measurements at the unknown sensor node can be expressed as [26]

$$P_r(d)[dBm] = P_0(d_0)[dBm] - 10\eta \log\left(\frac{d}{d_0}\right) + \chi_\sigma \quad (9)$$

where $d_0$ denotes the reference distance from the transmitter and $\eta$ represents the path loss exponent. $X_\sigma$ is zero mean Gaussian random variable with standard deviation $\sigma$. Unbiased estimate of true distance using Maximum likelihood estimate is given by

$$\tilde{d}_{ij} = d_0 \left(\frac{P_{ij}}{P_0(d_0)}\right)^{\frac{-1}{\eta}} \exp\left(\frac{-\log(10)\sigma^2}{20\eta^2}\right) \quad (10)$$

RSS a low cost measurement and it depends upon the wireless channel. It may yield large error because of the requirement of an accurate propagation model.

Semidefinite Programming Based Localization. RSS based method called semi-definite programming (SDP) is described in [27, 28], for noisy distance measurements as an approach to localization. The time measurement is expressed as

$$t_i = \frac{1}{v}\|X_i - Y\| + t_0 + w_i, \forall i = 1, \ldots, |S| \quad (11)$$

where $v$ and $t_0$ denote the speed of signal and unknown reference time. $w_i$ is Gaussian distributed with zero mean and $\sigma^2$ variance.

$$\hat{Y} = \underset{Y, t_0}{\mathrm{argmin}} \sum_{i=1}^{|S|} (t_i - \frac{1}{v}\|X_i - Y\| - t_0)^2 \quad (12)$$

SDP formulation of TDOA-based localization problem can be succinctly recast as

Minimize $\tau$
Subject to
$$-\tau < \frac{1}{\tau^2} \mathrm{Trace}\left(\begin{bmatrix} I & Y \\ Y^T & Y_s \end{bmatrix} \begin{bmatrix} X_i X_i^T & -X_i \\ -X_i^T & 1 \end{bmatrix}\right)$$
$$- \mathrm{Trace}\left(\begin{bmatrix} 1 & \frac{d_r}{v} \\ \frac{d_r}{v} & \frac{d_s}{v^2} \end{bmatrix} \begin{bmatrix} d_{ir}^2 & d_{ir} \\ d_{ir} & 1 \end{bmatrix}\right) < \tau \quad (13)$$
$$\begin{bmatrix} I & Y \\ Y^T & Y_s \end{bmatrix} \geq 0$$
$$\begin{bmatrix} 1 & d_r \\ d_r & d_s \end{bmatrix} \geq 0,$$

. The approach is less precise in estimating the location of static sensor nodes.

*MDS Based Localization*. In [29, 30], an attempt to reduce the error in estimated location for static nodes is discussed. The algorithm based on the multidimensional scaling (MDS) [31, 32], also computes the location of unknown nodes, given the set of distances between each pair of nodes. Finding the squared distance matrix.

$$D = \begin{bmatrix} 0 & d_{12}^2 & \cdots & d_{1|\mathcal{N}|}^2 \\ d_{21}^2 & 0 & \cdots & d_{2|\mathcal{N}|}^2 \\ \vdots & \vdots & \vdots & \vdots \\ d_{|\mathcal{N}|1}^2 & d_{|\mathcal{N}|2}^2 & \cdots & 0 \end{bmatrix}$$

Apply double centering matrix to D

$$D_{\mathrm{dc}} = JDJ \quad (14)$$

where,

$$J = I_{|\mathcal{N}| \times |\mathcal{N}|} - \frac{1}{|\mathcal{N}|}.1.1^T \quad (15)$$

where, 1 is the vector of all ones. Apply Singular Value Decomposition (SVD) on $D_{\mathrm{dc}}$

$$D_{\mathrm{dc}} = QAQ^T \quad (16)$$

where $A$ is the diagonal matrix having each element as eigen value and $Q$ a matrix corresponding to the eigen vector. Choosing the first 2 (For 2D) eigen value and corresponding eigen vector

$$D_{\mathrm{new}} = Q_{\mathrm{new}} A_{\mathrm{new}} Q_{\mathrm{new}}^T \quad (17)$$

Transform the relative coordinates system

$$\tilde{X} = Q_{\mathrm{new}} A_{\mathrm{new}}^{\frac{1}{2}} \quad (18)$$

to absolute coordinates system using anchor nodes.

The MDS-based localization method is energy efficient, as it only requires initial distance information. It also provides good initial coordinates for other optimization method. This method requires the set of anchors for converting the node location to actual coordinate system.

*Channel Impulse Response based Fingerprinting Localization.* Channel impulse response based fingerprinting localization [33] provides good accuracy compared to RSS based method. The mapping is carried out using Gaussian kernel. The received observation CIR vector at $i^{th}$ node can be expressed as

$$\mathbf{H_i} = \begin{bmatrix} h_{i1} & h_{i2} & \cdots & h_{i|\mathcal{S}|} \end{bmatrix}_{|\mathcal{S}| \times L}^T \quad (19)$$

$h_{ij}$ denotes the observation vector at $i^{th}$ node due to $j^{th}$ source of $L$ samples length.

$$\tilde{X} = \frac{1}{|\mathcal{N}|} \sum_{i=1}^{|\mathcal{N}|} \gamma_i X_i \qquad (20)$$

where,

$$\gamma_i = \frac{1}{\sqrt{(2\pi)^{|\mathcal{N}|}|\Sigma|}} \exp(-\frac{1}{2}(\hat{H} - H_i)\Sigma^{-1}(\hat{H} - H_i)) \qquad (21)$$

and sample covariance matrix

$$\Sigma = \frac{1}{|\mathcal{N}|} \sum_{i=1}^{|\mathcal{N}|} H_i.H_i^T \qquad (22)$$

### B. Range-free Localization Methods

Range-free methods use connectivity information and yields coarser location estimates than range-based methods. Range free localization methods like Approximate Point in Triangle Test (APIT) [34], DV-HOP [35, 36], Centroid algorithm [37], Monte-Carlo localization [38, 39], Closest point based method [40], Assumption based coordinates method [41] and Amorphous method [42] have been extensively dealt with in literature. Range-free localization methods are based on distance-hop or geometric configuration of sensor nodes [43, 2, 39, 38, 34]. Range free technique can be further grouped into two sub-categories [38] namely local techniques and hop-counting methods [42, 35].

*a) Approximate Point in Triangle Test (APIT) [34]*

Few sensor nodes that are equipped with high-powered transmitters are called anchors. The unknown node can obtain its location based on the information from the anchors. If there exists a direction of movement such that N is closer or further away to $A_1$, $A_2$, and $A_3$ simultaneously, then N is outside of $\Delta$ $A_1A_2A_3$ as shown in Figure 2. Otherwise, N is inside $\Delta$ $A_1A_2A_3$. The APIT is listed in Algorithm 1 and Figure 3. APIT is a cost-effective approach and it has good localization accuracy, even in the presence of irregular radio pattern, and with the random node deployment.

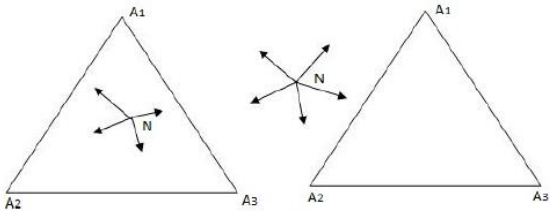

Figure 2: An illustration of point inside and outside of a triangle in APIT

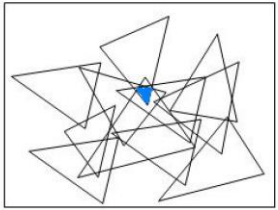

Figure 3: Illustration of the intersection of overlapping region in APIT, shown in Blue color

*b) Centroid-based Localization*

The centroid-based localization method [44] for sufficiently large number of anchor nodes is given by

$$\begin{bmatrix} \hat{x} \\ \hat{y} \end{bmatrix} = \begin{bmatrix} \frac{\sum_{i=1}^{|\mathcal{S}|} \alpha_i x_i}{\sum_{i=1}^{|\mathcal{S}|} \alpha_i} \\ \frac{\sum_{i=1}^{|\mathcal{S}|} \alpha_i y_i}{\sum_{i=1}^{|\mathcal{S}|} \alpha_i} \end{bmatrix} \qquad (23)$$

Where $[\hat{x} \ \hat{y}]^T$ is an unknown sensor node. $[x_i \ y_i]^T$ denotes the $i^{th}$ anchor and $\alpha_i$ represents the confidence of $i^{th}$ anchor communication. It is the most basic localization technique.

---
**Algorithm 1: APIT Localization**
Input: Location of $N$ Anchor
Output: Estimated location of node
1  Area $\leftarrow \phi$
2  for *each* $\Delta_i \in {}^NC_3$ do
3     if $\Delta_i == TRUE$ then
4        Area = Area $\cup \Delta_i$
5     if *accuracy*(Area) == *good* then
6        Break
7  Estimated Location = Centroid($\cap \Delta_i \in$ Area)
8  return *Estimated location*

---

*c) Monte-Carlo Localization*

The Monte-Carlo localization [38, 39, 129, 133] improves the localization accuracy by efficiently drawing the samples from the constrained area. The overlapped region of communication ranges of all neighboring anchors is called the anchor box. On the other hand, intersection of anchor box and range of node movement (based on the mobility of the sensor node) is called the sample box. The sample box further confines the area and it reduces the computational complexity, while filtering out the bad samples as shown in Figure 5. Those samples are subsequently utilized for obtaining the unknown sensor node. The steps involved for Monte-Carlo localization is listed in Figure 4.
Input:

$$\text{Observation} = (\text{AnchorID}_j, \ x_j, \ y_j), \forall j = 1, \ldots, |\mathcal{N}| \qquad (24)$$

Anchor box building is described by

$$x_{\min}^t = \max(x_j - R_{\text{anchor}}, \ 0) \qquad (25a)$$
$$x_{\max}^t = \min(x_j + R_{\text{anchor}}, \ x_{\text{net}}) \qquad (25b)$$
$$y_{\min}^t = \max(y_j - R_{\text{anchor}}, \ 0) \qquad (25c)$$
$$x_{\max}^t = \min(y_j + R_{\text{anchor}}, \ y_{\text{net}}) \qquad (25d)$$

where the dimensions of the network is assumed as $[0, x_{net}] \times [0, y_{net}]$. The radio communication range of the anchor is denoted by R. $[x^t_{min}, x^t_{max}] \times [y^t_{min}, x^t_{max}]$ represents the anchor box at time *t*. In this work, min and max are the minimum and maximum operators respectively. Similarly, sample box building is described by

$$x_{\min}^i = \max(x_{\min}^t, \ x_i^{t-1} - v_{\max}, \ 0) \qquad (26a)$$
$$x_{\max}^i = \min(x_{\max}^t, \ x_i^{t-1} + v_{\max}, \ x_{\text{net}}) \qquad (26b)$$
$$y_{\min}^i = \max(y_{\min}^t, \ y_i^{t-1} - v_{\max}, \ 0) \qquad (26c)$$
$$y_{\max}^i = \min(y_{\max}^t, \ y_i^{t-1} + v_{\max}, \ y_{\text{net}}) \qquad (26d)$$

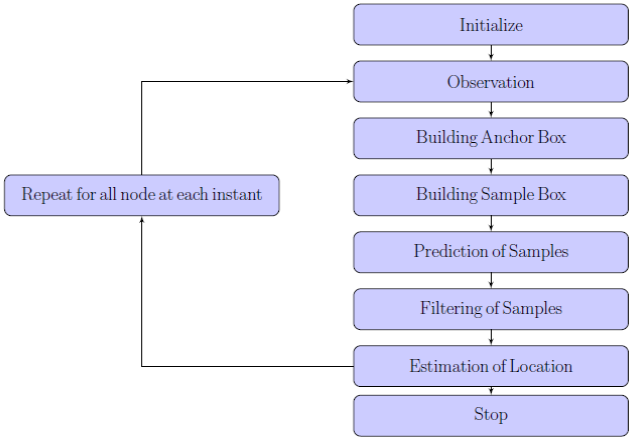

Figure 4: Monte-Carlo localization algorithm

### d) DV-Hop based Localization

The DV-HOP localization method is based on the average hop distance between two sensor nodes. It does not utilize the absolute distance between two neighboring sensor node and consequently gives coarser location of node than range-based methods. The DV-HOP based localization method [36] is given by

$$\text{Hop}_{\text{average}} = \frac{\sum_{i,j}\sqrt{(x_i - x_j)^2 + (y_i - y_j)^2}}{\sum_{i,j} h_{i,j}} \quad (27)$$

where, $(x_i, y_i)$ = Coordinates of the $i^{th}$ anchor, and $h_{i,j}$ = Hop count from $i^{th}$ node to $j^{th}$ node.

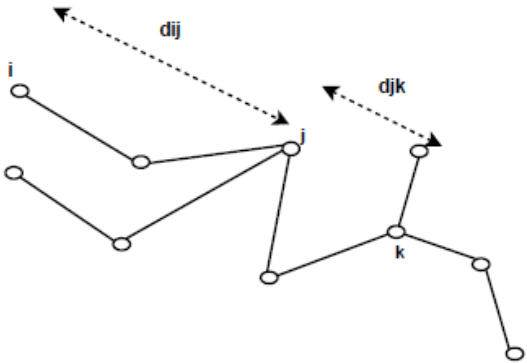

Figure 6: Illustration of the DV-HOP localization

It may be noted that computing the accurate distance between two sensor nodes is highly erroneous especially in indoor scenario or a large sensor network.

### e) Closest Point based Localization

The closest point localization [40] provides a low cost solution, easily implementable, low energy consumption and low resolution of location estimate. The closest point localization method is employed for obtaining the cardinal location information. The coarse location of the unknown sensor node is decided based on the following conditions:
• Node receiving maximum power closest to which anchor?
• Whether node is in communication range of $i^{th}$ anchor?
• Direction in which nth node receives maximum energy?
• Near-far information between pair of nodes.

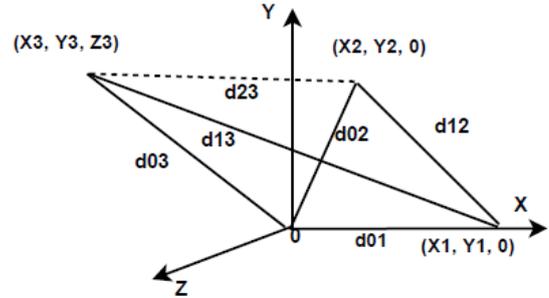

Figure 7: An illustration of node placement in ABC algorithm

### f) Assumption Based Coordinates (ABC) Localization Method

Assumption Based Coordinates (ABC) [41] is a distributed localization method. Initially, it is difficult to get the range measurements from more than three anchors (for 2D coordinate system). Therefore, a map based on the neighboring ranges is generated at the beginning of the network operations. The Assumption Based Coordinates (ABC) localization method [41] is explained in Figure 7. The following geometry based expressions for the computation of node locations are given by

$$x_2 = \frac{d_{01}^2 + d_{02}^2 - d_{12}^2}{2d_{01}} \quad (28)$$

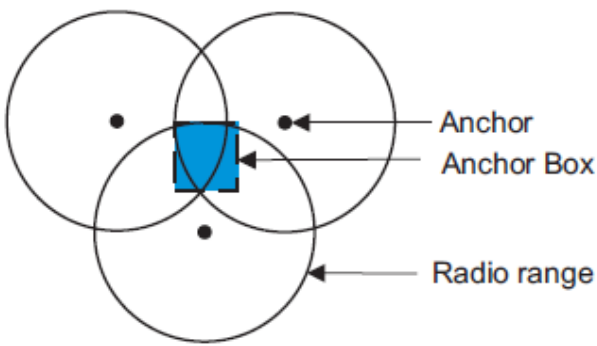

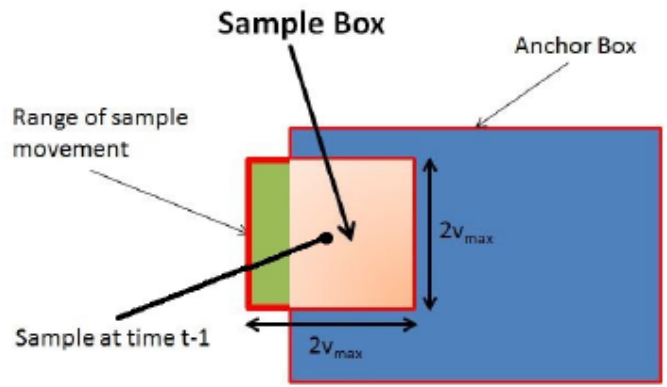

Figure 5: Illustration of a (a) building the anchor box and (b) building the sample box [45]

$$y_2 = \sqrt{d_{02}^2 - x_2^2} \qquad (29)$$

$$x_3 = \frac{d_{01}^2 + d_{03}^2 - d_{13}^2}{2d_{01}} \qquad (30)$$

$$y_3 = \frac{d_{03}^2 - d_{23}^2 + x_2^2 + y_2^2 - 2x_2x_3}{2y_2} \qquad (31)$$

$$z_3 = \sqrt{d_{03}^2 - x_3^2 - y_3^2} \qquad (32)$$

Range-based and range-free localization methods can be implemented using various state-of-the-art methods. Some of them are described in ensuing section.

### III. OVERVIEW OF IMPLEMENTATION METHODS

In this section, a brief overview of various implementation methods for node localization relevant to this localization and tracking is provided. Various machine learning method used in the implementation of localization are first listed. Subsequently, implementation of localization algorithm based on computational architecture is discussed. A brief review of fingerprinting and multi-sensor data fusion based localization method is also provided [125].

#### A. Machine Learning Based Methods

The performance of all the aforementioned methods like RSS, TDOA, TOA and AOA methods degrade under harsh weather conditions [46]. On the other hand, machine learning techniques such as neural networks, and Hidden Markov Model require a large training data. But they have added advantage of performing well under noisy conditions. When using Hidden Markov Model (HMM), the system is modeled as a Markov process with hidden states [47]. In the context of localization, the hidden states refer to the range corresponding to the given observed sequence. This helps in estimating the distance between an anchor and a node from which the signal is received. This distance information obtained from HMM model is then further used for node localization. In [8], stochastic RSS-map model and RSS channel model are discussed. However, these methods are not robust to environmental changes. Because of time varying nature of wireless channel, RSS [8] method performs badly. RSS fluctuates due to wall, furniture, and people movements in indoor NLOS conditions. Obstacles such as building, and forests contribute to creation of outdoor NLOS conditions. In this context, Bayesian network called semi supervised hidden Markov model has the advantage. NLOS identification and mitigation using least square support vector machine and Gaussian processes for Wi-Fi RSS data is proposed in [48]. Semi-supervised Laplacian regularized least squares method and HMM based RSS-EKF (Extended Kalman Filter) method using RSS are described in [49, 50] respectively.

#### B. Centralized and Distributed Methods

Broadly, localization algorithms can be implemented with a centralized or a distributed architecture. Localization based on centralized architecture such as location estimation using convex radial constraints [51] and multidimensional scaling (MDS-Map) [52, 29], and simulated annealing [53] requires a powerful base station and increases in terms of memory and power requirement. Each node participates in localizing the rest of the nodes, in this class of co-operative localization. Hence, requirement of anchors may be relaxed and can be named as anchor-free localization algorithms [37, 54, 55]. On the other hand, distributed architecture utilizes a large number of resource constrained wireless sensor nodes. Distributed approach can also speed-up the localization process. In this method, anchors within vicinity of an unknown node are utilized for location estimation [56]. The position of the anchor nodes needs to be known in advance either through Global Positioning System (GPS) or any other location aware devices. At-least three anchor nodes are required in two-dimensional space for removing rotation and translation ambiguity of the network. However, distributed localization algorithm under-performs in the presence of a large number of sensor nodes and may also result in loss of information. Network security is also an issue with the distributed architecture. Hence, a decentralized architecture is a better choice with respect to optimizing the overall limitations.

#### C. Multi-Sensor Data Fusion Based Methods [57]

In order to enhance the accuracy of a localization system, a fusion based approach is widely used. Wi-Fi based slam [58, 59, 60, 61] fuses RSS and motion sensor data for simultaneous building a map and locating a user. The RAVEL, radio and vision enhanced localization system, which fuses Wi-Fi with visual data is explored in [62]. Further, the organic landmark maps utilize the unique identifiable signatures within the building [63, 64]. These signatures then correct the dead-reckoning error for enhanced accuracy in this unsupervised localization method. In contrast to these methods, the algorithm that utilizes low cost multi-sensor data such as radio, acoustic, visible light and fused signal performs better in terms of localization error.

#### D. Fingerprinting Based Methods [57]

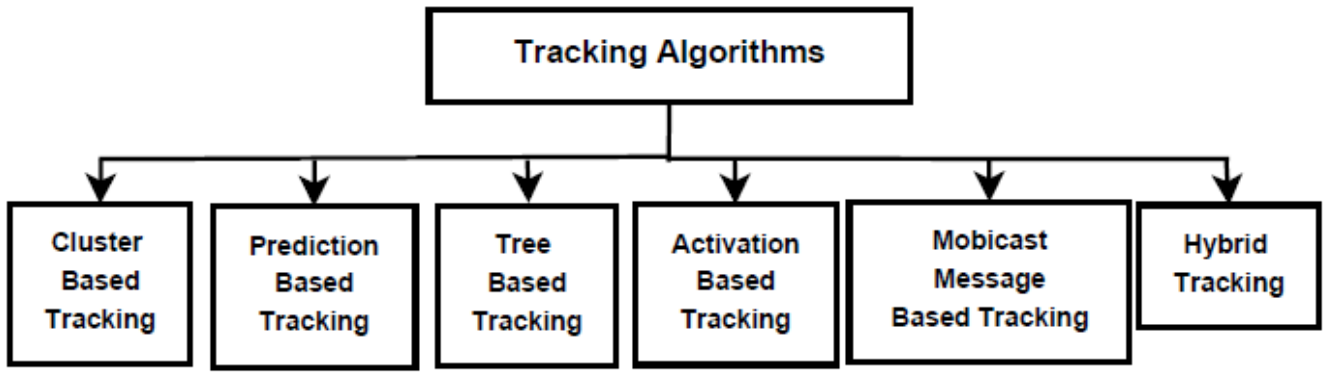

Figure 8: Broad categorization of tracking algorithms

Standard fingerprinting based localization methods available in the literature are Radar [65], PlaceLab [66], and Horus [67]. These methods comprise primarily two phases: training and testing [126]. During training phase, radio map is generated at each of the fingerprint location. Subsequently, a location is chosen corresponding to the minimum error between test RSS data and fingerprint RSS. [68, 69]. This method is able to cope with heterogeneous devices during training and testing phases. A method described in [68] utilizes ratio of RSS, whereas, [69] uses relative RSS data for coarse-grained localization due to which performance deteriorates [70]. Notably, these fingerprinting based localization methods provide good location accuracy at the expense of heavy training during radio map construction. Classification of tracking algorithms is described in the ensuing section.

## IV. Broad Classification of Tracking Methods

Tracking of mobile node [2] is carried out generally using a Kalman filter [71], an extended Kalman filter [72] or a particle filter [73, 74] in current state of-art methods. The mobile sensor node accounts for all possible motion scenarios unlike a static node [127]. Mobile nodes can be localized using range free techniques. Applications in this context include vehicular tracking in a city [75]. Networks which consist of mobile nodes also employ Monte Carlo localization techniques [38, 39].

In literature, several approaches on mobile node tracking [76, 77, 78, 79, 80] have been discussed. The tracking algorithms can be classified as shown in Figure 8. Several tracking techniques are combination of two or more methods, which is categorized as hybrid tracking methods [81, 82, 83, 84, 85, 86, 87, 88, 89, 90, 91].

### A. Cluster-based Tracking Methods

There are three types of clustered-based approaches, namely static, dynamic and spatio-temporal. In [92], three cluster based approaches, adaptive head, static head, and selective static head are proposed. The adaptive head [92] is the most energy efficient approach while static and selective static heads algorithms are chosen for rapidly moving target. Cluster management strategies, such as the proactive and the reactive cluster management are presented to handle energy and the longevity constraints of the sensor nodes [93]. A dynamic clustering method that optimizes the information utility of data for a given cost of communication and computation is explored in [94]. However, this energy efficient approach is applicable to single target only. An energy aware target detection and localization strategy for cluster-based wireless sensor network is presented in [95]. This probabilistic localization method is not scalable and limited to a single target. A decentralized, lightweight, dynamic clustering algorithm for target tracking is proposed in [88]. A cluster head is chosen for the received acoustic signal above a set threshold. The cluster head collaborates with the neighboring sensors for the exchange of information. In [96], a tracking approach that uses dynamic mechanism to deal with different ways of missing targets is presented. DELTA [97] tracks a constant moving target dynamically by building groups and electing group leaders using flashlight. An attempt to reduce the number of the number of nodes involved in communication for tracking and consequently decreasing energy consumption is presented in [98]. In [84], a classification algorithm for target tracking is developed. It estimates the velocity to reduce uncertainty in the movement of the target and thus tracking error. However, this method is applicable only for single target. A hybrid static/dynamic clustering method called continuous object detection and tracking algorithm (CODA) is described in [99]. It utilizes cluster head to estimate the boundary profile and communicates this to base-station.

#### a) Prediction-based Tracking Methods

The prediction based tracking is one of the subclass of clustered-based tracking. In this context, an energy efficient prediction based mobile node tracking method is discussed in [78, 100, 87]. This method is energy efficient since sensors are in sleep mode most of the time and avoid long range transmission to the base-station. The architecture of this prediction-based energy saving scheme (PES) comprises of three blocks, namely, prediction, wake up and recovery. Only the sensor node that tracks a mobile target is active, while remaining sensor nodes stay in a low-power mode. Hence the sensing and micro-control unit are deactivated. Sensor nodes monitor and send the information of mobile target to the base-station. The prediction from base-station is sent back to the sensor node, when the target enters its range. The sensor node continues to monitor and track the object, if the prediction is consistent. Otherwise, the target is declared missing and a recovery mechanism is activated. However, the work is limited for the case of moving patterns of the mobile target.

In [101], dual prediction based reporting (DPR) for mobile target tracking is discussed. Both sensor node and base-station predict the future location of target. Multi-hop communication does not take place in case of consistent prediction. The DPR comprises of two subsystem; location and prediction. The location model determines the granularity of the target location state and prediction model estimate targets' future movement. In [91], distributed processing for multiple target tracking is presented. Distributed architecture is preferred over centralized architecture for energy efficiency in a wireless sensor network. A linear predictor (DPT) that utilizes previous two locations to predict the next location is proposed in [89]. The method is distributed and is scalable while minimizing the energy consumption of the network. However, reduction of miss rate, relationship between energy consumption and location accuracy, and extension for mobile sensors are the open problems here. An optimal tracking interval for a given predictive tracking of the target is proposed in [102]. The energy aware tracking system with error handling ability is discussed in [103]. This method considers the case of error avoidance and error correction besides tracking. The method also handles sudden direction change of moving target and transmission failure. However, varying speed of mobile target is not taken into the account.

*B. Tree-based Tracking Methods*

A scalable tracking using networked sensors (STUN) [104] tracks a large number of targets using hierarchical organization. In [105], a DCTC method to detect and track a target by pruning and expanding a convoy tree. An initial convoy tree is created, when target is detected. The neighboring sensor nodes send the data to the root node and the tree is reconfigured once the target moves out. In [106, 107], an approach for target tracking that ensures reasonable localization accuracy, low energy dissipation, and low complexity is proposed. However, this work is not suitable for multiple intruding objects. In-network data processing for object tracking is carried out using update cost and query cost efficiently in [108, 109]. A query based efficient target tracking that continuously reports the tracking information to a mobile user is presented in [110]. A voronoi-based target tracking is explored in [83], which establishes non-uniform coverage to detect the target using non-redundant sensor nodes

*C. Activation-based Tracking Methods*

There are four kinds of activation based tracking algorithms [111, 112, 113, 114, 115]. In Naive activation method, each node communicates to the base station to achieve good tracking accuracy. However, it suffers from high energy consumption. In random activation method, each node is active with a fixed probability and nodes are chosen randomly for tracking. In selective activation method, a small subset of nodes closer to next predicted location of the target are chosen. The duty-cycle activation method turns on sensor nodes in certain periods of time and is used jointly with other activation based methods

*D. Mobicast-based Tracking Methods*

It is a spatio-temporal method that forwards messages (location and time) to nodes in some geographic zone. The messages in HVE-mobicast method are distributed in two phases namely, cluster-to-cluster and cluster-to-node phases. The method is energy efficient because it takes into account different moving speed and directions in a geographic zone [116, 86, 117]. Compactness information of the network is utilized to guarantees reliable spatio-temporal message delivery [118, 119, 90]. A space-time clustering method for narrowband target localization and tracking is explored in [120]. It is based on the closest point of approach (CPA) of a target to the sensor nodes. However it requires significantly larger number of sensor nodes and thus requires a large amount of energy.

V. EXPERIMENTAL SETUP FOR LOCALIZATION AND TRACKING

In this section, we review the leading experimental kits that can be utilized for localization and tracking purpose. However, there are several experimental kits available for research and development [121] such as National Instruments (NI) WSN, Crossbow motes [122, 123], SensWiz, hand-held devices, Dust networks kit, and TKN wireless indoor sensor network testbed (TWIST). Note that, most of the kits have similar on-board sensors, but, differs primarily in cost and specifications. The experimental kits may be chosen based on the application-specific requirements.

*A. National Instruments Wireless Sensor Nodes*

National Instruments (NI) WSN - 3202, and 3212 nodes and NI 9792 gateway can be used to create the ad-hoc networks. These NI WSN nodes have capability to measure the link quality between sensor node and gateway. The distance is inversely proportional to the link quality. A general WSN node can also be configured as a router to extend the network coverage. The outdoor communication range of NI wireless sensor node is 300 m, whereas, the indoor range is 100 m. The communication among sensor nodes is based on the IEEE 802.15.4 protocol. The sensor data is first sensed by the sensor node and then transmitted to the gateway. Subsequently, it is interfaced with LabVIEW on the central system. LabVIEW is an integrated development environment that is used to create and code the localization system. Hence, it extends its applicability for various applications. Note that, the communication range of NI wireless sensor node is large and having high cost. Additionally, LabVIEW is not an open-source software. The NI wireless sensor node can measure different signal modalities like light, pressure, strain, sound, ultrasonic, and infrared. Those multi-sensor data can be subsequently utilized for localization and tracking purpose.

*B. Crossbow Motes*

Crossbow motes MTS310 sensor boards, XM2110 IRIS board and MIB520 USB mote interface are employed to create the networks. It is initially developed by University of California, Berkeley. Tiny-OS platform is used for the setup, however, sensor data can be exported in the Matlab. The physical dimensions of the Crossbow motes are small compared to NI wireless sensor nodes and it is relatively cheaper. IEEE 802.15.4 is used for communication protocols and it works in 2.4 to 2.48 GHz, a globally compatible ISM band. The data rate is 250 kbps. Additionally, it utilizes direct sequence spread spectrum that is robust to RF interference and gives data security. However, its radio communication range is small, typically 15 m - 20 m. It may noted that the Crossbow mote also comprises of various on-board sensors like light, acoustic, temperature, magnetometer, dual-axis accelerometer, received signal strength indicator (RSSI), health packet, and battery level. Since, the wireless channel is time-variant in nature and hence, the packet loss is also varying with the time. Thus, the Crossbow motes selfconfigure itself.

*C. SensWiz Networks Kit*

SensWiz is developed by Dreamajax Technology, India. Similar to NI and Crossbow motes, SensWiz also operates on 2.4 GHz with IEEE 802.15.4 communication protocols. It also has on-board sensors like light, temperature, humidity and tri-axis accelerometer axis. It gives flexibility to users to modify/write their own programs using simple programming interface.

*a) Hand-Held Devices*

In this modern age, several hand-held devices such as smart-phone, tablet, laptop, smart wrist watch, and goggles can act as a sensor node. Similar to the others experimental kits, these devices can acquire different signal modalities like visual, radio, humidity, and temperature. There are several applications (app) available on the play-store. The app can be used for the collection of sensor data for subsequent operations. The location computation may be performed on the go, or in the central system. In the case of centralized architecture, sensor data may be sent via blue tooth (for short range), internet (for long range) or universal serial bus (USB).

CONCLUSION

Localization accuracy of the range based methods depend upon reliability of radio propagation model, whereas, fingerprinting localization yields good accuracy at the expense of extensive offline training. The low-cost multi-sensor data fusion based localization provides good localization accuracy. The sensor node tracking is mostly applicable for the mobile sensor networks. It requires additional network coordination among all the sensor nodes. The application of mobile sensor networks includes disaster management, emergency response and military surveillance. The paper provides a comprehensive survey of localization and tracking methods and also the experimental set-ups for the specific application such as localization and tracking. This survey paper is targeted for the audience who wish to pursue the research in this area.